# Target Fishing: A Single-Label or Multi-Label Problem?

Avid M. Afzal[1], Hamse Y. Mussa [1§], Richard E. Turner[2], Andreas Bender [1§] and Robert C. Glen[1]

[1]Centre for Molecular Informatics, Department of Chemistry, University of Cambridge, Lensfield Road, Cambridge CB2 1EW, United Kingdom

[2] Computational and Biological Learning Lab, Department of Engineering, University of Cambridge, Trumpington Street, Cambridge CB2 1PZ, United Kingdom

[§]Corresponding author(s)

Email addresses:

    AMA: maa76@cam.ac.uk

    HYM: mussax021@gmail.com

    RT: ret26@cam.ac.uk

    AB: ab454@cam.ac.uk

    RCG: rcg28@cam.ac.uk

# Abstract

**Background:** According to Cobanoglu *et al* and Murphy, it is now widely acknowledged that the single target paradigm (one protein/target, one disease, one drug) that has been the dominant premise in drug development in the recent past is untenable. More often than not, a drug-like compound (ligand) can be promiscuous – that is, it can interact with more than one target protein.

In recent years, in *in silico* target prediction methods the promiscuity issue has been approached computationally in different ways. In this study we confine attention to the so–called ligand-based target prediction machine learning approaches, commonly referred to as *target-fishing*.

With a few exceptions, the *target-fishing* approaches that are currently ubiquitous in cheminformatics literature can be essentially viewed as single-label multi-classification schemes; these approaches inherently bank on the single target paradigm assumption that a ligand can (somehow) home in on one specific target. In order to address the ligand promiscuity issue, one might be able to cast *target-fishing* as a multi-label multi-class classification problem.

For illustrative and comparison purposes, single-label and multi-label Naïve Bayes classification models (denoted here by SMM and MMM, respectively) for *target-fishing* were implemented. The models were constructed and tested on 65,587 compounds/ligands and 308 targets retrieved from the ChEMBL17 database.

**Results:** SMM and MMM performed differently. According to Recall and Precision evaluation metrics employed to compare the target prediction performance of the two models for 16,344 test compounds/ligands, the MMM model returned recall and precision values of 0.8058 and 0.6622, respectively; the corresponding recall and precision values yielded by the SMM model were 0.7805 and 0.7596, respectively. However, at a significance level of 0.05 and one degree of freedom McNemar's test performed on the target predication results returned by SMM and MMM for the 16,344 test ligands gave a p-value $<$ 7. $\times 10^{-5}$ ($\chi^2$ value of 15.656), in favour of the MMM approach: The MMM model correctly predicted the targets of 262 test compounds, which the SMM model wrongly predicted; and the SMM model correctly predicted the targets of 178 test compounds, which the MMM model wrongly predicted.

**Conclusions:** The target prediction results obtained in this study lend support (albeit statistically) to the argument against the single target paradigm. The results also indicate that multi-label multi-class approaches are more robust and apt than the ubiquitous single-label multi-class schemes when it comes to the application of ligand-based classifiers to *target-fishing*.

# Background

It is now widely acknowledged that the single target paradigm (i.e. one protein/target, one disease, one drug) that has been the dominant premise in drug development in recent past is untenable as both drug-like compound (ligand) and target protein can be promiscuous [1][2]. More often than not, a ligand can simultaneously interact with multiple proteins in a human cell; this observation can also be true with target proteins [2][3]. For example, according to Mestres [4], there is on average $6 - 7$ annotated targets per drug in DrugBank [5]. It is, therefore, important that ligand (and protein) promiscuity is taken into consideration when developing *in silico* target protein prediction models. In this regard, significant efforts have been made in recent years to take into account the promiscuity issue when devising *in silico* target protein prediction models [1]–[3][6]–[9] (and references there in). The state-of-the-art methods that consider ligand (and protein) promiscuity when predicting target proteins can be broadly divided into three categories namely ligand-based [1][3][6][7][10][11], target-structure-based [1][3][6][8], and ligand-target-pair-based [1][3][6][9]. In this study we confine attention to ligand-based machine learning approaches, commonly referred to as *target-fishing*.

The central idea that constitutes the nub of the ligand-based machine learning approach is that a new ligand sharing enough structural similarity to a set of reference ligands annotated against known target proteins has a high probability of showing activity against the predefined target proteins [6] (and references therein).

The *target-fishing* approach began to appear in the cheminformatics literature over the last decade and a half [10]–[21]. According to Rognan [6], the *target-fishing* methods all share three basic components:-- using a given data set comprising a set of reference ligands, a set of target proteins and a bipartite activity relation between the targets and ligands in the two sets, a model is constructed such that for a new ligand the model returns the appropriate targets against which this ligand shows activity.

As far as we are aware, at the time of writing, the ligand-based machine learning approaches – with few exceptions (see the *Previous Work* Section) – utilised in cheminformatics explicitly or implicitly assume that the target proteins against which the reference ligands are annotated are mutually exclusive [3][6][10][11][15][17][22]–[24] (and references therein). It is assumed a ligand can (somehow) home in on one single protein in the midst of the multitude of proteins in a human cell, which is the very assumption deemed questionable [1][2][4], see above. In machine learning (and also in statistics), this type of ligand-based target predicting approach can be viewed as a single-label multi-class classification problem, *vide infra*. In contrast, as in this work, one might be able to take into account ligand promiscuity by casting the ligand-based target prediction task/approach as a multi-label multi-class classification problem. That is, the relevant target proteins for a certain ligand need not be mutually exclusive. This will be described in detail in the *Methods* Section.

In any event, in the light of the discussion in the preceding paragraphs the machine learning ligand-based target predicting approach (*target-fishing*) is basically a ligand-based classification problem [3][6][22]–[24][31], whereby a (machine learning) classifier is utilised to predict potential target protein(s) for a given ligand. Thus, developing an accurate, computationally efficient and conceptually appropriate ligand-based classifier is an important research topic in cheminformatics. To this end, the essence of devising an efficient ligand-based classification model can amount to developing a mathematical algorithm that "learns" the chemical structure-biological activity relationships (if any) from given set of (reference) ligand chemical structures, a predefined set of target proteins and a bipartite activity relation between the reference ligands and targets. Once the learning phase of the model/classifier building is completed, for a new compound the resultant classifier is expected to accurately predict relevant target proteins (in the preselected set of target proteins) against which the new compound may show biological activity.

The ligand chemical structure is usually represented as a "vector" (descriptor/feature vector) whose elements, ideally, constitute the salient characteristics of the ligand for its interaction with potential target protein(s). There are a plethora of chemical structure representation schemes that have been suggested over the years [25][26]. Simply one cannot predicate that a given representation of a chemical structure can capture all the subtleties intrinsic to a particular chemical structure of the ligand, which might be crucial for the biological effect that a ligand could induce on the relevant target proteins. Another source of uncertainty is the certitude that measurements of observable biological effects (and subsequently databases based on these observations) are inevitably noisy [27][28]; this uncertainty can, in turn, introduce another layer of uncertainty in relating the chemical structure of the ligand with its observable activity against a target protein. It is, therefore, desirable to develop a ligand-based classification approach that takes into account these uncertainties. This deems a probabilistic classifier the ideal candidate for the task [19][24][29][30][33]–[35].

In more concrete terms, a ligand-based classifier can be viewed as an algorithm that maps a ligand descriptor vector $\mathbf{x}_j$ to a predefined protein target(s) often referred to as classes/labels. Henceforth, all the target labels are collectively denoted by $L = \{l_1, \ldots, l_k, \ldots, l_{|L|}\}$. Usually $\mathbf{x}_j$ is viewed as a "vector" defined on an $m$-dimensional descriptor space $\chi$, where $\mathbf{x}_j = \{x_{j1}, \ldots, x_{ji}, \ldots, x_{jm}\}$. As described before, the elements $x_{ji}$ are assumed to represent the "relevant" chemical structure descriptors/properties of ligand $j$ in relation to the targets. These descriptors can assume real or discrete values. In the present work $x_{ji}$ are binary, representing the absence or presence of a chemical atom environment descriptors in the ligand.

A tacit assumption that is typically made is that one has access to a representative data set $D$ that adequately captures the bipartite activity relation between the target proteins and reference ligand chemical structures: $D = \{(\mathbf{x}_j, Y_j), j = 1, 2, \ldots, N\}$ denoting the $N$ available data points, where $\mathbf{x}_j \in \chi$ represents ligand $j$ and $Y_j$ refers to the set of targets against which ligand $\mathbf{x}_j$ is known to be active.

Given $D$, the classification task amounts to "learning" or estimating a function (if one exists):

$$f: \chi \to L \quad (1)$$

which not only correctly associates the known label(s) $Y_j$ with their appropriate ligand $\mathbf{x}_j$, but also predicts the correct label(s) for a new ligand that is not included in $D$. In effect, our main task is to come up with a model that elucidates or captures the unknown underlying process that might have generated the observed phenomena, i.e. the available data set $D$, in the first place.

In **Eq. 1** the function can denote a ligand-based deterministic or probabilistic classifier [33]–[37]. In the present work, attention has been confined to Naïve Bayes classifiers, which are probabilistic. In this case, both $\mathbf{x}_j$ and $Y_j$ are random variables, but for notational simplicity in this work both $Y_j$ and $\mathbf{x}_j$ denote both the random variables and the values they may assume. Furthermore, unless stated otherwise, the index $j$ in $\mathbf{x}_j, Y_j$ and $x_{ji}$ and the indices $j$ and $k$ in $l_{kj}$ are omitted for notational clarity, where $l_{kj}$ refers to label $k$ for compound $j$.

In the pattern recognition literature [38]–[45], when $|Y| = 1$, a classification model is referred to as a single-label classifier; but when $|Y| \geq 2$, the classification model is referred to as a multi-label classifier. Furthermore, a classification problem can also be referred to as a binary classification problem if $|L| = 2$ and a multi-class classification problem when $|L| > 2$. Thus, a multi-class classification task can be categorised as a multi-label multi-class classification or a single-label multi-class classification problem. For an extended and detailed account of the multi-label multi-class classification topic the reader may consult refs. [38]–[45].

As discussed in the *Previous Work* Section (see below), to our best knowledge the current *target-fishing* approaches employed in cheminformatics (with a few exceptions) rely on the assumption that a given ligand can only interact with one target protein, i.e. $|Y| = 1$. Thus, a ligand-based target predicting approach, probabilistic or not, can be viewed as a single-label classification model. Nonetheless $|L|$ separate single-label classifiers (one classifier per class) are constructed and utilised in order to predict potential multiple targets for a given ligand [10][11][13][15] [17][22][23]. These |L| classifiers can be induced binary (one–vs–all) [10][12][15], or multi-class classifiers "proper" [11][13][17][22][23]. In this work we employed the latter classifiers as they are more apt and robust than the one–vs–all classifiers [23] in multi-class classification problems. In any event, the high probability of a ligand interacting with more than one target protein in nature [1]–[4], [47]–[51] – i.e. $|Y| \geq 2$ – may render the single-label classification approach questionable as a *target-fishing* scheme.

In the light of our earlier discussion one may consider a ligand-based target prediction approach as a multi-label multi-class classification task when $|Y| \geq 2$ and $|L| > 2$. Since target proteins/classes are not necessarily mutually exclusive in the case of $|Y| \geq 2$, a single multi-label multi-class ligand-based classifier should, ideally, be able to capture the underlying association (if any) between the chemical structures of the ligand and the set of labels $Y \subseteq L$ denoting potential target proteins for this ligand. Thus, this single multi-label classifier should, in principle, be able to predict the relevant target protein(s) for a given ligand. Having said that, nothing stops one from constructing |L| individual induced binary (but "pseudo single-label") classifiers for the same purpose, providing that the given training data set *D* is appropriately transformed (for a detailed account of training data set transformation in the multi-label classification context, see ref. [39]). It should be noted that there are subtle but crucial differences between the induced binary classifiers employed in single-label ligand-based models described earlier and the induced binary classifiers (termed "pseudo single-label" here) employed in multi-label classification settings. This issue is briefly commented on in the *Methods* Section, but for a more detailed description, see ref. [38]. In our present study, |L| individual induced binary "pseudo single-label"

classifiers were constructed and employed, where the data transformation scheme utilised was *binary relevance* [39].

Arguably classification approaches based upon Naïve Bayes constitute the bulk of the probabilistic classification models for *target-fishing* [10][12][13][15][19] (and references therein). For this reason, we concentrated on this particularly popular ligand-based classification model. The popularity of the Naïve-Bayes as a *target-fishing* tool can be probably attributed to the fact that building non-Naïve Bayes multi-class classifiers (be probabilistic or not) can become conceptually intricate or computationally demanding, or both [10][11][17][18][19][22]–[24][35]–[37]. The Naïve Bayes approach is: (1) probabilistic; (2) favourably scalable with $m$, $L$, and $N$, where $m, L$ and $N$ are as defined before; (3) computationally simple to implement; and (4) known to yield respectable classification results, despite the flimsiness of the rationale upon which the algorithm is based (that is, descriptors for a ligand are conditionally independent of each other given the class label). It is these characteristics that give the application of Naïve Bayes based *target-fishing* approaches an edge over other classification algorithms also employed for this purpose [19] (and references therein).

**Previous Work**

For more recent developments on *target-fishing* approaches, we refer the reader to refs. [1][3] and [6]. To our knowledge, there were no research papers, at the time of writing, regarding the topic of comparing single-label and multi-label multi-class Naïve Bayes classifications for *target-fishing*. Michielan *et al*. [20] employed multi-label multi-class classification to classify cytochrome p450 substrates. The authors employed multi-label multi-class classification models based on SVM, $M_L$K-Nearest-Neighbour, and Neural Network on a data set of 580 cytochrome p450 substrates and seven isoforms. Hristozov *et al.*[21], also employing SVM, Neural Network, and $M_L$K-Nearest-Neighbour methods [42], looked into classifying sesquiterpene lactones into seven tribes from the plant family Asteraceae. The two research groups compared the performance of single-label and multi-label models, and cautiously noted that multi-class classifiers based on the multi-label concept outperformed, or performed just as well as their corresponding single-label multi-class classifiers. However, their work did not feature the subject matter here: Naïve Bayes algorithms; besides, compared to ours their studies covered only seven targets. Wale and Karypis employed multi-label ligand-based classification methods [16]. Unlike our study, the nub and the main thrust of Wale and Karypis's work were about comparing how different multi-label ligand-based classifiers perform on classifying multi-label bioactivity data sets. Similarly Kawai *et al.*'s study [29] was confined to the analyses of the performance of a multi-label ligand-based SVM classifier; the single-label aspect did not feature in their work, nor did single-label and multi-label Naïve Bayes algorithms.

Closely following studies in text mining [52], we implemented and studied a ligand-based Naïve Bayes multi-label multi-class classification model (MMM) for *target-fishing*. We compared this classifier with a single-label multi-class ligand-based Naïve Bayes classification model (SMM) designed for the same purpose. Both classification models were built and tested on a bioactivity data set extracted from the ChEMBL17 database [53], which comprised 308 protein target classes and 65,587 compounds.

In the following and preceding sections the words "ligand" and "compound" are used interchangeably. Also the terms "class", "activity", "label", "target" and "target protein" are employed interchangeably. Single-label and multi-label compounds mean that a compound is non-promiscuous and promiscuous respectively. A single-label data set refers to a data set containing only single-label compounds, whereas a multi-label data set refers a dataset comprising promiscuous compounds.

## Materials and Methods

**Data set**

In order to construct the Naïve Bayes models, we used the ChEMBL17 database, which comprises more than 1 million annotated compounds and more than 10 million bioactivity records covering 9,000 targets. The data set used for this study was a subset of ChEMBL17, which consisted of 65,587 unique compounds covering 308 human targets giving a total of 93,281 ligand-target pairs. Structures with reported activities (IC50/ki/kd/EC50) equal or better than 1μM and confidence scores of 8 or 9 against human protein targets were selected. Although this bioactivity value represented highly potent compounds, given the increase in the size of ChEMBL17 database, it represented a sensible trade-off between biological activity and coverage of the chemical space. Only protein classes that contained between 120 and 720 data points were selected to ensure that the data set was balanced.

**Table 1** summarises our ChEMBL17 data set. The table shows that 83.1% of the total compounds were annotated against only one target protein, while the remaining 16.9% of compounds were annotated against two or more protein targets.

**Table 2** and **Figure 1a** depict the distribution of target proteins in different protein families. The majority of target proteins are categorised as enzymes and membrane receptors, with enzymes representing 67.8% of all the protein targets/classes in our ChEMBL17 data set, and membrane receptors constituting 23% of it. **Figure 1b** depicts the distributions of the enzyme classes. A significant proportion of the enzyme families in our data set consisted of the Kinase and Protease classes, with 54% and 15%, respectively. 7TM1 receptors constitute the bulk (89%) of all the membrane receptor classes in our data set (see **Figure 1c**).

Since there were a significant number of multi-label compounds (more than one-sixth of the total number of compounds) in our data set, we believe, this was a suitable data set for testing the hypothesis described in the *Background* Section.

**Compound Descriptors**

Compounds were standardized prior to fingerprint generation by ChemAxon's Standardizer [54] using the options "Remove Fragments", "Neutralize", "Remove

Explicit Hydrogen" and "Tautomerize". Extended Connectivity Fingerprints (ECFP) were employed to describe compound structures [55]–[58]. ChemAxon's Java API [54] was utilized to generate fixed-length ECFP_4 binary fingerprints with a length of 1,024 bits.

**Methods**

In this section we briefly describe the single-label and multi-label multi-class Naïve Bayes algorithms that were employed in this study.

**Naïve Bayes**

According to the Naïve Bayes assumption, the descriptors $\{x_1, x_2, x_3, \ldots, x_m\}$ constituting the elements of the descriptor vector **x** representing the ligand are assumed conditionally independent given the class label $l$ [19]. In this setting, a choice of $f$ (in **Eq. 1**) can be the class posterior probability $p(l \mid \mathbf{x})$, where $p(l \mid \mathbf{x})$ can be expressed as [19]

$$p(l|\mathbf{x}) = \frac{\prod_{i=1}^{m} p(x_i \mid l) \, P(l)}{p(\mathbf{x})} \quad (2)$$

where $P(l)$ refers to the a priori probability of the class label $l$. This term represents one's state of knowledge about the class label before obtaining the data for the ligands. The term $p(x_i \mid l)$ denotes the class (label) conditional probability for $x_i$, and $p(\mathbf{x})$ is as defined below; **x**, $|L|$ and $m$ are as described before. In this study, $x_i$ is binary – i.e. $x_i \in \{0,1\}$. Comparatively, it is a simple affair to estimate $P(l)$. Thus, in practice, the estimation of $p(l|\mathbf{x})$ reduces to the estimation of $\prod_{i=1}^{m} p(x_i \mid l)$, i.e. the $p(x_i \mid l)$'s.

**Single-label Multi-class Naïve Bayes**

In the single-label multi-class Naïve Bayes model employed in this work, where $|Y| = 1$ and $|L| > 2$, $p(\mathbf{x})$ was expressed as $p(\mathbf{x}) = \sum_{l=1}^{|L|} \prod_{i=1}^{m} p(x_i \mid l) \ P(l)$.

The class conditional probability $p(x_i \mid l)$ was estimated as

$$p(x_i \mid l) = \frac{1 + n_{il}^+}{2 + n_l} \qquad (3)$$

where $n_{il}^+$ denotes the number of times that the $i\,th$ descriptor $x_i$ assumes the value 1 in class $l$ and $n_l$ is the number of instances in the training set belonging to class $l$. The a priori probability of each class $P(l)$ was estimated as

$$P(l) = \frac{n_l}{N} \qquad (4)$$

where $N$ denotes the total number of single-label training data points. ($\mathbf{x}$, $|L|$ and $m$ are as described before.)

One classifier was built for each target protein $l$ using **Eqs. 2 – 4** and the compounds in the training data set that were annotated against this target only. For predicting potential target proteins for a new compound, SMM computes $|L|$ class/target posterior probability values and outputs the class with the highest posterior probability value.

**Multi-label Multi-class Naïve Bayes**

The multi-label multi-class Naïve Bayes algorithm, with $|Y| \geq 2$ and $|L| > 2$, was implemented based on Wei et al. [52], where a *binary relevance* transformation [39] was utilised. However, any other appropriate transformation of the training data set could have been employed [39]. Wei *et al.*'s approach is briefly described below for completeness. For a detailed account and more erudite description of what transforming the training data set entails in the multi-label context, the reader is referred to ref. [39].

Using **Eqs. 2 – 4** and a *binary relevance* transformation $|L|$ induced binary (but "pseudo single-label") classifiers, $H_l: \chi \rightarrow \{l, \neg l\}$, were constructed – one for each unique label $l$ in the set $L$. This means, to construct the $|L|$ binary classifiers, the original data set $D$ was transformed into $|L|$ data sets $D_l$, where each one of them contains all the instances of the original data set $D$. Each compound is labelled active if it is labelled $l$ in the original data set and otherwise labelled negative by the class label $\neg l$. Note that in the multi-label Naïve Bayes case, unlike the single-label Naïve Bayes discussed in the *Background* Section, a compound in the training set needs not be a single-label compound. That is, the induced binary classifiers are not strictly single-label classifiers in the multi-label case – hence, the attribute "pseudo single-label".

To predict the appropriate class labels (potential target proteins) for a new test compound **x**, the multi-label multi-class classification scheme outputs the union of the labels predicted by the $|L|$ classifiers, $Z$:

$$Z = \bigcup_{l \in L} \{l: H_l(\mathbf{x}) \geq p_{threshold}\} \quad (5)$$

where $H_l(\mathbf{x})$ denotes $p(l|\mathbf{x})$ for compounds **x**. Here $p(l|\mathbf{x})$ was computed via **Eq. 2**; $p(\mathbf{x})$ was given by $p(\mathbf{x}) = \prod_{i=1}^{m} p(x_i | l) P(l) + \prod_{i=1}^{m} p(x_i | \neg l) P(\neg l)$; $p(l) = \frac{n_l}{N}$

, whereas $p(\neg l) = 1 - p(l)$; and $p(x_i \mid l)$ and $p(x_i \mid \neg l)$ were estimated by using **Eq. 3**.

The parameter $p_{threshold}$ was tuned/optimised by using cross-validation based on the Recall – Precision scheme.

## Computational Details

**Model Evaluation Schemes**

In the multi-label multi-class classification case, a class prediction made by a multi-label multi-class model (MMM) can be fully correct, partially correct or fully wrong. Hence, the evaluation schemes for MMM are more complicated than those employed for evaluating the generalisation ability of a single-label multi-class model (SMM), whose prediction can only be fully correct or fully wrong. In order to make the comparison of MMM and SMM as equitable as possible, a rejection threshold value was not specified. Instead, in both MMM and SMM, a class prediction is arbitrarily assigned to one class when two or more classes are equally predicted.

We employed Recall and Precision evaluation schemes.

For MMM, recall and precision evaluation measures based on ref. [44] are widely employed in the machine learning literature; we followed suit:

$$precision = \frac{1}{|T|} \sum_{j=1}^{|T|} \frac{|Y_j \cap Z_j|}{|Z_j|} \quad (6)$$

$$recall = \frac{1}{|T|} \sum_{j=1}^{|T|} \frac{|Y_j \cap Z_j|}{|Y_j|} \qquad (7)$$

where $T$ denotes the multi-label test data set which has $|T|$ examples $(\mathbf{x}_j, Y_j), j = 1 \dots |T|, Y_j \subseteq L$; and $Z_j = \cup_{l \in L}\{l: H_l(\mathbf{x}_j) \geq p_{threshold}\}$ represents the set of labels to which $\mathbf{x}_j$ is predicted to belong.

However, in the case of SMM, recall and precision values are computed (per class) as

$$precision = \frac{TP}{TP + FP} \qquad (8)$$

$$recall = \frac{TP}{TP + FN} \qquad (9)$$

where "$TP$" denotes the number of compounds that the model assigns to their actual target, (say) target A; "$FN$" refers to the number of compounds annotated against target A, but assigned to other targets, whereas "$FP$" represents the number of compounds whose associated target was wrongly predicted to be target A.

Evaluating the generalisation ability of SMM – using **Eqs. 8** and **9** – was straightforward. However, the same cannot be said about MMM because in this case the predictions can be partially correct, fully correct and fully wrong. Thus, to make the comparison of the classification performance of the two models on the single-label data set at hand as equitable as possible, only the predicted class label in the top position of the predicted set of class labels $Z_j$ for compound $\mathbf{x}_j$ is considered as the predicted class label when computing $|Y_j \cap Z_j|$ in **Eqs. 6** and **7**. It should be noted that, while this approach puts the "recalls" in **Eq. 7** and **Eq. 9** on equal footings, it heavily penalises the precision value in the MMM case as the denominators in **Eqs. 6** and **8** indicate.

## Model Construction and Testing

Our ChEMBL17 data set was randomly split into two portions – 70% of it as a training set $S$, and from the remaining 30%, only single label compounds as a test set.

Using the 70% ChEMBL17 dataset allotted to training, the multi-label multiclass classification and single-label multiclass classification models based on the Naïve Bayes concept were generated, see *Methods*.

The multi-label multiclass classification model (MMM) was built on the available training data set $S$. The single-label multiclass classification model (SMM) was built only on single-label training data set. This specific training data set was generated from $S$ by simply associating each compound with only one of its targets, where the target with the highest measured bioactivity was retained.

To compare the classification performance of MMM and SMM (on the remaining 30% of our ChEMBL17 data set) we utilised the two evaluation schemes described in the previous section.

The Recall and Precision metrics were used to compare the performance of the two models on the single-label data set. In SMM there was no parameter to estimate. However, in MMM, the optimal choice of the $p_{threshold}$ value (described in the *Methods* Section) had to be estimated, which was achieved via 5-fold cross validation on the training set. The best threshold value for each fold was computed and the mean value of these threshold values was considered as the "optimal" $p_{threshold}$ value. For all the results given in the following section $p_{threshold}$ was set to 0.999.

## Results and Discussion

It is worthy of note that the classification predictions (and the subsequent analyses) presented in this study were retrospective in the sense that the predicted targets were known beforehand.

**Classification Performance**

The two classification models, MMM and SMM, were tested on predicting the appropriate targets for 16,344 test compounds distributed over 308 target proteins. Columns 2 and 3 in **Table 3** demonstrate the target prediction performance of the two models for the test data set: SMM returned recall and precision values of 0.7805 and 0.7596 (Column 2), respectively; the corresponding recall and precision values yielded by MMM were 0.8058 and 0.6622 (Column 3), respectively.

SMM yielded a better precision value for the test data set, but this could be attributed to the Recall and Precision evaluation metrics employed. As described in the *Methods* Section, this evaluation scheme heavily penalises (see the denominators in **Eqs. 6** and **8**) the precision value returned by MMM. The scheme, however, puts the recalls returned by MMM (**Eq. 7**) and SMM (**Eq. 9**) on equal footings. In this case the MMM classifier returned better recall value than the SMM classifier. Since the Recall – Precision metric was inconclusive, further analyses were performed. These analyses revealed that MMM and SMM statistically behave differently, based on our dataset For example, at a significance level of 0.05 and with one degree of freedom, McNemar's test performed on the MMM and SMM target prediction results for the test set yielded a p-value $< 7. \times 10^{-5}$ and $\chi^2$ values of 15.657 – in favour of the MMM approach; because the targets for 178 test compounds were wrongly predicted by the MMM model (but correctly predicted by the SMM model), whereas the targets for 262 test compounds were correctly predicted by the MMM model (but wrongly predicted by the SMM model), see **Table 4.**

These analyses suggest that MMM statistically generalises better than SMM based on the training data sets utilised. Thus, one could argue that the *target-fishing* results yielded by our multi-label and single-label models certainly – albeit statistically – lend support to the argument against the single-target paradigm and *target-fishing* methods that are based on this paradigm.

## Conclusion

In this work two *in silico* ligand-based target prediction models – single-label multi-class and multi-label multi-class Naïve Bayes classifiers – were constructed and tested on a large data set of bioactivity data extracted from the ChEMBL17 database. Statistically, the multi-label multi-class classification model (MMM) significantly outperformed its corresponding single-label multi-class classification model (SMM) on predicting the appropriate target proteins for 16,344 ChEMBL17 test compounds.

At a significance level of 0.05 and with one degree of freedom, McNemar's test performed on the MMM and SMM target prediction results for the test compounds yielded a p-value $< 7. \times 10^{-5}$ and $\chi^2$ value of 15.657, in favour of MMM.

The target prediction results obtained are in line with the hypothesis set out within this study, i.e., it is not appropriate, nor is it adequate to universally employ single-label multi-class ligand-based classification approaches for *target-fishing*. Thus, based on the data sets utilised in this work, one may conclude that out of the two classification approaches (SMM and MMM) tested, the multi-label multi-class model (MMM) is robust and more apt (and should be utilised) for ligand-based *target-fishing* purposes – the subject matter in this study.

## List of Abbreviations

SMM: Single-label Multi-class Model.

MMM: Multi-label Multi-class Model.

SEA: Similarity Ensemble Approach.

PASS: Prediction of Activity Spectra for Substances.

## Competing interests

The authors declare that they have no competing interests.

## Authors' contributions

AMA generated the ChEMBL datasets, implemented and evaluated the algorithms presented in this work. He also wrote a major part of the manuscript. HYM contributed to the main theme on which the worked was performed, the writing up of the manuscript and to the implementation of the algorithm. RT contributed to the analysis of the results; AB and RCG ensured that the pharmaceutical aspect of the work was rationally valid. All authors contributed to revising the final draft of the manuscript.

## Acknowledgements

AMA would like to thank the Centre for Molecular Informatics for its support. HYM, AK, AB and RCG acknowledge support by Unilever. We also we would like to thank Dr John Mitchell for insightful comments.

# Figures

**Figure 1 - Protein target distribution in the ChEMBL17 data set.**

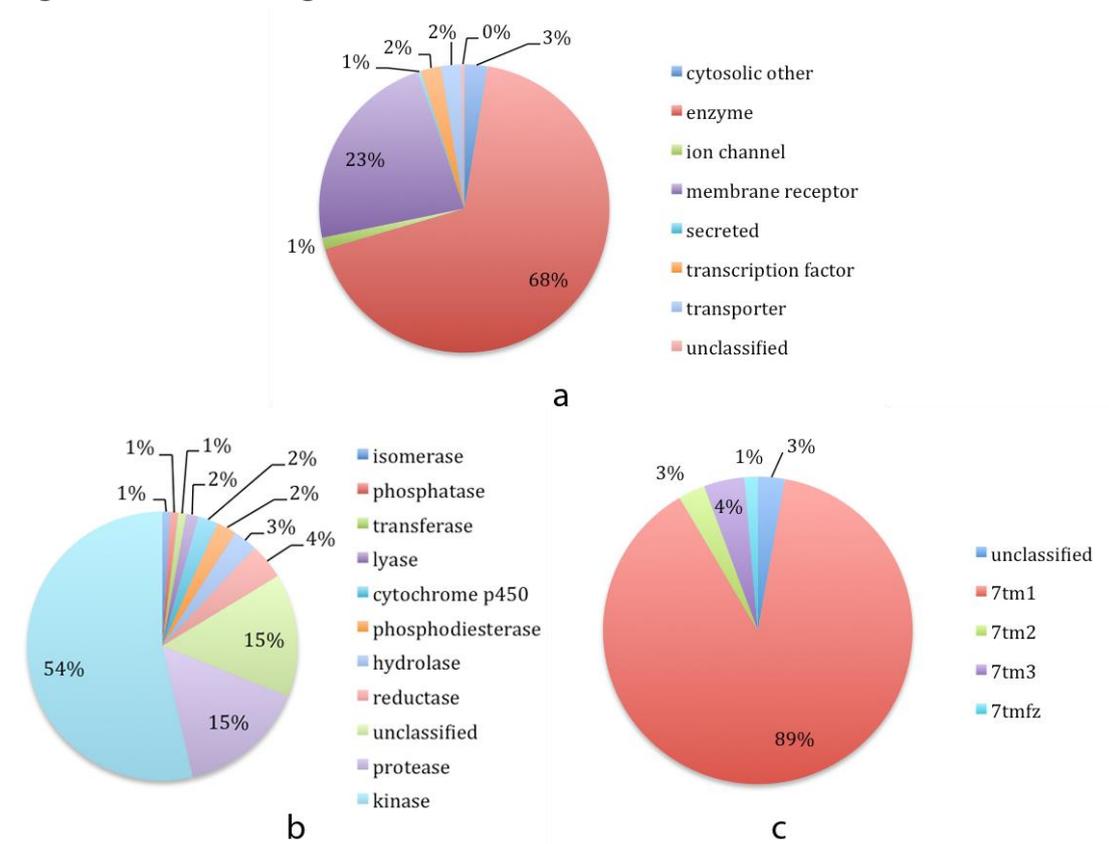

**Figure 1a:** Protein target distribution among protein families in the ChEMBL17 data set. **Figure 1b:** The distribution of protein targets in enzyme families. **Figure 1 c:** The distribution of protein targets in membrane receptor families.

# Tables

**Table 1 – Distribution of the compounds and their associated protein targets in our ChEMBL17 dataset.**

| Number of Annotated Targets | 1 | 2 | 3 | 4 | 5 | 6 | 7 | 8 | 9 | ≥ 10 |
|---|---|---|---|---|---|---|---|---|---|---|
| Number of Compounds | 54,563 | 7,937 | 1,571 | 321 | 240 | 191 | 132 | 60 | 42 | 530 |
| % of Total Number of Compounds | 83.1% | 12.1% | 2.39% | 0.49% | 0.36% | 0.29% | 0.2% | 0.09% | 0.06% | 0.8% |

The table shows that 83.1% of the total compounds were annotated against only one target protein, while the remaining 16.9% of compounds were annotated against two or more protein targets; just over one-sixth of our ChEMBL17 dataset comprises multi-label ligands.

**Table 2 - Distribution of target proteins in different protein families in our ChEMBL17 dataset.**

|  | Enzyme | membrane receptor | transcription factor | ion channel | transporter | cytosolic other | Unclassified | secreted |
|---|---|---|---|---|---|---|---|---|
| Number of Classes | 209 | 71 | 7 | 4 | 7 | 8 | 1 | 1 |
| % of Total Classes | 67.85% | 23.05% | 2.27% | 1.29% | 2.27% | 2.59% | 0.32% | 0.32% |

90.90% of the protein targets are enzymes and membrane receptors, with enzymes representing 67.85% of all the protein targets, and membrane receptors constituting 23.05%.

**Table 3 – Recall and Precision values returned by MMM and SMM on predicting the target proteins for the 16,344 test compounds.**

| Model | SMM | MMM |
|---|---|---|
| Recall | 0.7805 | 0.8058 |
| Precision | 0.7596 | 0.6622 |

**Table 4 – The McNemar's test for the 16,344 test compounds.**

|  | Number of test compounds |
|---|---|
| MMM predicted correctly while SMM predicted incorrectly | 262 |
| MMM predicted incorrectly while SMM predicted correctly | 178 |
| McNemar's test Result | 15.6568 (7.594e-5) |

The McNemar's test at a significance level of 0.05 and with one degree of freedom, on the MMM and SMM target prediction results for the test set yielded a p-value < 7. $\times 10^{-5}$ and $\chi^2$ value of 15.657.